\documentclass[pra,twocolumn,showpacs,floatfix]{revtex4}
\usepackage{graphicx}

\newcommand{\beq}{\begin{equation}}
\newcommand{\eeq}{\end{equation}}
\newcommand{\beqa}{\begin{eqnarray}}
\newcommand{\eeqa}{\end{eqnarray}}
\newcommand{\ba}{\begin{array}}
\newcommand{\ea}{\end{array}}

\begin{document}

\title{Nonlinear Schr\"odinger equation for a  superfluid Fermi
gas from BCS to Bose crossover}
\author{S. K. Adhikari
}
\affiliation{
Instituto de F\'{\i}sica Te\'orica, UNESP - S\~ao Paulo State
University, 01.405-900 S\~ao Paulo, S\~ao Paulo, Brazil}

\begin{abstract}
We introduce a  quasi-analytic nonlinear Schr\"odinger equation with
beyond mean-field corrections
to
describe the dynamics
of a zero-temperature dilute superfluid Fermi gas in the crossover from
the weak-coupling
Bardeen-Cooper-Schrieffer (BCS)
regime, where $k_F|a| \ll 1$ with $a$ the
s-wave scattering length and $k_F$ the Fermi momentum,
through
the unitarity limit,  $k_Fa  \to  \pm \infty$, to the Bose
regime where $k_Fa >0$.
The energy of our model
is parametrized using the known asymptotic behavior in the BCS, Bose,
and the unitarity limits and is in excellent agreement with accurate
Green function Monte Carlo calculations. The model generates good 
results for frequencies of  collective breathing oscillation of a 
trapped Fermi 
superfluid.

\end{abstract}

\maketitle

The crossover from a Bardeen-Cooper-Schrieffer
(BCS)
Fermi superfluid at zero temperature for weak coupling to a Bose
condensate of dimers  \cite{1}
has been an intense area of research 
(both
experimental \cite{excross,excross1} and
theoretical \cite{TH,2,th1,th3,MS,KZ,PS})  after 
the realization of  a 
BCS to Bose crossover (BBC)
 in a trapped dilute Fermi superfluid
near a Feshbach resonance.
This  allows
to change the system from the BCS regime, with
small  negative $a$  ($k_F|a|\ll 1$), through
the unitarity regime
of divergent $a$  ($k_Fa\to  \pm \infty$), to the Bose regime of
dimers with positive $a$, with $k_F$ the Fermi momentum.
The limiting behavior of the system in the
weak-coupling  ($k_F|a|\ll
1$) \cite{HY,heisel,baker,th2,lhy} and
unitarity limits  for both positive and negative
$a$ are well known \cite{heisel,baker,BB}. Accurate information about
the  BBC dynamics has
recently been available  from  numerical fixed-node Green's
function Monte Carlo (GFMC) calculations \cite{th1,th3} at zero
temperature. 
There 
is
also mean-field BCS (MFBCS) calculation of the same \cite{2}.

A quasi-analytical model for
the BBC
problem, generating a  the Ginzburg-Landau (GL) type equation for
fermions 
for
$a<0$ and a Gross-Pitaevskii (GP) type  equation for dimers for
$a>0$, both including beyond mean-field effects, so as to be valid in
the
unitarity limit of divergent $a$,  should be useful.
We propose such a model in three dimensions,
called the BBC
model
(BBM),
for the crossover  problem of
a dilute trapped Fermi superfluid at zero temperature
using the known theoretical solution in the weak-coupling and unitarity
limits in both the BCS and Bose regimes.

We consider a dilute superfluid Fermi gas of $N$ spin-half atoms  of
mass $m$, atomic scattering length $a$,  
and density $n$ with
singlet pairing due to an attractive atomic interaction. 
We assume, for a dilute gas,   that the
results are universal determined solely by the scattering length $a$
and
independent of the detail of the interaction potential and its range.
We present  an analytical model for energy and bulk chemical potential
in the 
entire crossover region and hence derive a 
nonlinear equation for the superfluid Fermi gas. We present 
 results for radial and 
axial frequencies of collective oscillation   in a cigar-shaped trap and 
compare 
with GFMC and MFBCS calculations as well as experimental data.

At low densities, in the BCS regime
($k_F|a|\ll 1$), gaps are negligible  \cite{heisel,TH} and the total 
energy $E$ per
particle of the
superfluid Fermi gas
is given by \cite{HY,baker}
\beqa\label{a3}
\frac{E}{N}=\frac{3}{5}E_F\biggr[1+\frac{10}{9\pi}
k_Fa+\frac{4(11-2\ln
2)}{21\pi^2}(k_Fa)^2\nonumber \\+0.030467(k_Fa)^3-0.062013
(k_Fa)^4+...\biggr],
\eeqa
where $E_F\equiv \hbar^2k_F^2/(2m)=  An^{2/3}\equiv 
\frac{\hbar^2}{2m}(3\pi^2n)^{2/3}$ is the Fermi energy 
with $k_F=(3\pi^2n)^{1/3}$ \cite{TH}.

In the Bose regime ($a>0$),  the paired fermions form a weakly repulsive Bose gas of
dimers of mass $m'=2m$ and density $n'=n/2$ with a 
dimer-dimer scattering length $a'= 0.6a$, as predicted by Petrov
{\it et al.} \cite{petrov}. In this
regime, the energy of the system
is given by \cite{TH,lhy,th3}
\beqa
\frac{E}{N}+\frac{\varepsilon_B}{2}=\frac{3}{5}E_F
\left[\frac{5(k_Fa')  }{18\pi}
+\frac{64 (k_Fa')^{5/2} }
{27\sqrt{6\pi^5}}+...\right], \label{dimer}
\eeqa
with $a'=0.6a$ \cite{petrov},
where   $\varepsilon_B$ is the (positive) binding energy of the dimer.
 The dynamics of the dimer bosons is governed by the energy
$(E/N+\varepsilon_B/2)$ discounted for dimer binding.
The lowest-order term in Eqs. (\ref{a3}) and (\ref{dimer})
leads
 to the standard mean-field GL equation for Fermi
superfluid  and GP equation for dimer bosons, respectively,
 with
higher-order terms leading to beyond mean-field corrections due to
interaction.

The limiting energies (\ref{a3}) and (\ref{dimer}), valid for small
$|a|$, drastically fail as one approaches the unitarity limit
$a\to \pm \infty$. For a dilute Fermi superfluid, the unitarity limit is
supposed to be universal and the relevant energy scale is the energy of
the noninteracting Fermi gas $3E_F/5$    \cite{th2,cowell}, so that the
energy of the system is given by \cite{heisel,baker}
$
E/N= 3E_F\xi /5
$
 with $\xi$ a
universal factor.
By using Pad\`e approximant 
 Baker \cite{baker} estimated
 two values for   $\xi =0.326$ and 0.568, and
Heiselberg \cite{heisel}  obtained $\xi = 0.326$.
Later, from
experimental results
the following values for $\xi$ have been obtained
$\xi=0.51 \pm 0.04$ \cite{ex1},  0.7 \cite{ex2},    
$0.27^{+0.12}_{-0.09}$ \cite{ex3}, $0.41\pm 0.15$ \cite{tar}, $0.46\pm 0.05$ \cite{par}, 
and $0.46^{+0.05}_{-0.12}$ \cite{ste}.  The accurate
theoretical estimates for the factor $\xi$  are  $\xi=0.44 \pm 0.01$
\cite{th1} and
$0.437 \pm 0.009$
(GFMC)  \cite{th2}, and 0.59 (MFBCS) \cite{2}.
 Bulgac and Bertsch \cite{BB}
suggested the following approximate
behavior of energy per particle near the $(k_F|a|)^{-1}\to 0$ limit
\beqa \label{a4}
\frac{E}{N}= \frac{3}{5}E_F[\xi-\zeta (k_Fa)^{-1}-\frac{5}{3} v 
(k_Fa)^{-2}+...]
\eeqa
in both  BCS ($a<0$)
and Bose  ($a>0$)  regime with $\zeta\approx v\approx  1$.
In the Bose regime ($a>0$),
one should replace $\frac{E}{N}$ by 
$\frac{E}{N}+\frac{\varepsilon_B}{2}$. 


We
write a simple expression for energy $E/N$ combining the limiting
behaviors (\ref{a3}) and (\ref{a4}) in fermion  ($a<0$)
and   dimer
boson ($a>0$)
regime.
We suggest the following  expression for
fermion energy for $a<0$
\begin{eqnarray}
\frac{ E}{N}
= \frac{3}{5}E_F\left[
1+\frac{\frac{10k_Fa}{9\pi}}{1-\frac{10k_Fa}{9\pi(1-\xi)}}\right].
\label{a2}
\end{eqnarray}
This expression has the constant
$10/(9\pi)$ taken from limit (\ref{a3}) and the
constant $\xi$
from  limit
(\ref{a4}).
Expression (\ref{a2}) reproduces the  first two terms in
Eq. (\ref{a3}) exactly for small $|a|$.
It also reproduces the first term in  Eq. (\ref{a4})
for large $|a|$,
for the second term it yields $\zeta =0.9186$ in place of the
approximate value $\zeta \approx 1$

For $a>0$,  we use
the asymptotic behaviors of  energy for large and small $a$,
Eqs. (\ref{dimer}) and (\ref{a4}), to propose the following
expression for energy 
\beqa
\frac{E}{N}+\frac{\varepsilon_B}{2}=\frac{3}{5}E_F\left[
\frac{\frac{5}{18\pi}(k_Fa')+\frac{64}
{27\sqrt{6\pi^5}}
 (k_Fa')^{5/2}}{1+  \frac{64(a'/a)
(k_Fa')^{3/2}}{27\xi^2 \sqrt{6\pi^5}}+\frac{64(k_Fa')^{5/2}}
{27\xi \sqrt{6\pi^5}  }}\right].
\label{dimer2}
\eeqa
For large  $a'$,
by construction, expression (\ref{dimer2}) satisfies Eq. 
(\ref{a4}) with $\zeta =1$.
For small $a$  it reproduces the first term of expansion
(\ref{dimer}) exactly and the next terms  closely.

The only free parameter in model Eqs. (\ref{a2}) and
(\ref{dimer2}) (termed
BBM)
is the
universal
factor $\xi$, for which we use $\xi=0.44$  \cite{th1,th2,th3}.
The $k_Fa$ dependence of BBM (\ref{a2}) and (\ref{dimer2}) are
consistent with the behavior for small and large $k_F|a|$.
Kim and Zubarev \cite{KZ} 
considered  two different  [2/2] Pad\`e approximants to parametrize 
energy 
in the Fermi superfluid and Bose dimer regimes. Manini and Salasnich 
\cite{MS} considered  an energy function with arctan dependence on 
scattering length both in the  Fermi superfluid and Bose dimer regimes. 

\begin{figure}[tbp]
\begin{center}
{\includegraphics[width=.7\linewidth]{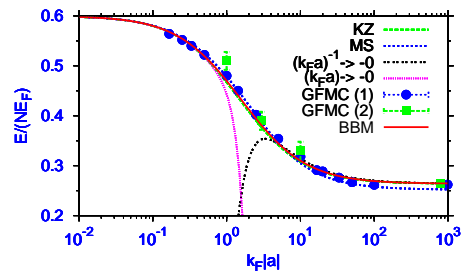}}
\end{center}
\caption{(Color online)
 $E/(NE_F)$ vs. $k_F|a|$  for $a<0$:
BBM $-$
Eq. (\ref{a2}); KZ  from Ref. \cite{KZ}; MS
from  Ref.  \cite{MS}; limiting value $(k_Fa)^{-1}\to -0$
of Eq. (\ref{a4});
limiting value $k_Fa\to -0$ of  Eq.
(\ref{a3}); GFMC (1)  \cite{th2}; GFMC (2)
 \cite{th1}.
}
\label{fg1}
\end{figure}

\begin{figure}[tbp]
\begin{center}
{\includegraphics[width=.7\linewidth]{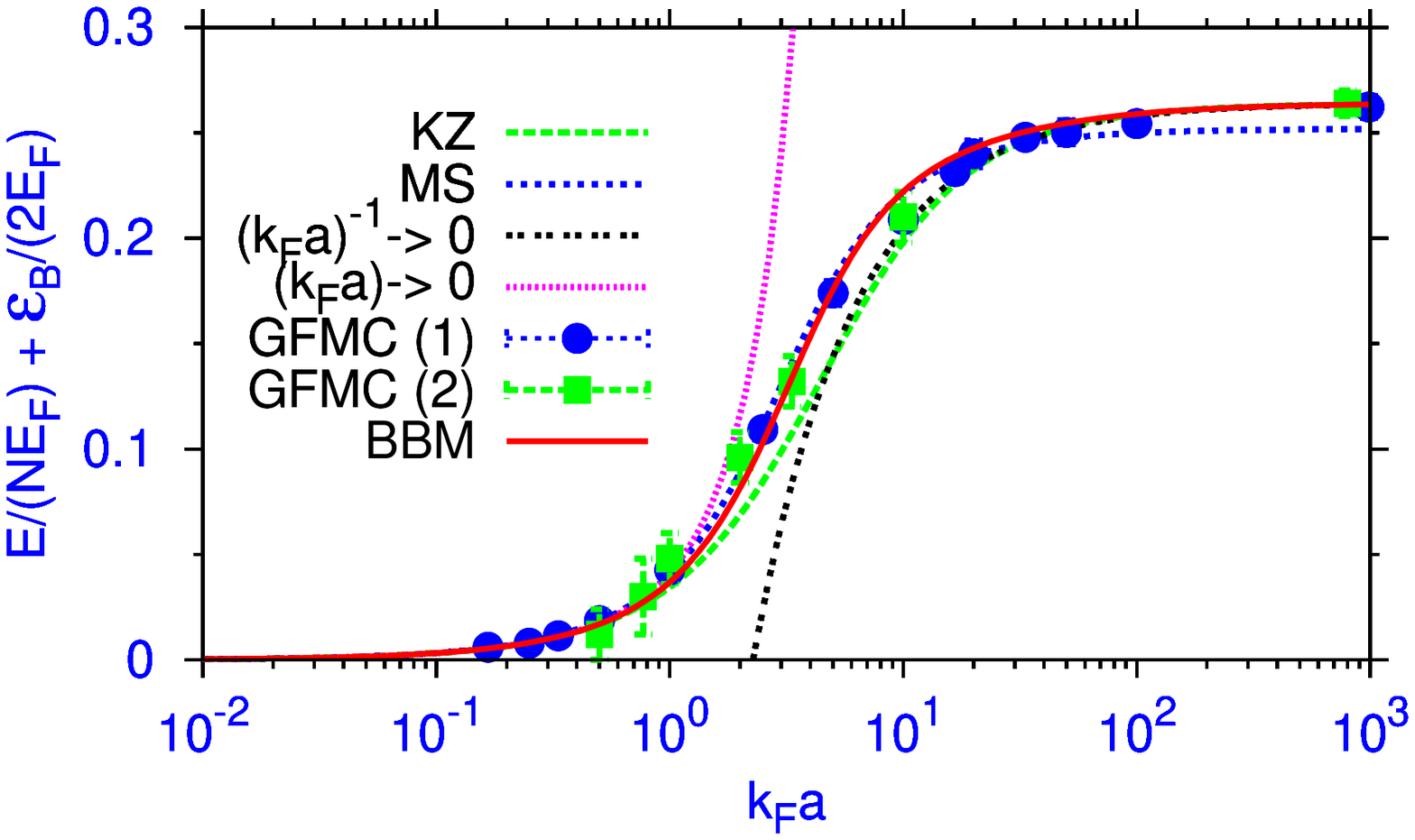}}
\end{center}
\caption{(Color online)
$E/(NE_F)+\epsilon_B/(2E_F)$ vs. $k_Fa$  for $a>0$:
BBM $-$
Eq. (\ref{dimer2}); KZ from Ref. \cite{KZ}; MS
from Ref.   \cite{MS}; limiting value $(k_Fa)^{-1}\to +0$
of Eq. (\ref{a4});
limiting value $k_Fa\to +0$ of  Eq.
(\ref{dimer}); GFMC (1)  \cite{th2}; GFMC (2)
 \cite{th1}.
}
\label{fg2}
\end{figure}

Next we
plot in  Fig. \ref{fg1}, for $a<0$,
the  energy from Eq. (\ref{a2}), in addition to those from  limits (\ref{a3})
and (\ref{a4}). We also plot the results of GFMC calculations  of
Refs.
\cite{th1,th2} and parametrizations of Refs. \cite{KZ,MS}.
Next
we plot, for $a>0$,  in
Fig. \ref{fg2}, the results for energy from Eq. (\ref{dimer2}),
the  asymptotic limits (\ref{dimer}) and
(\ref{a4}), the GFMC results of Refs. \cite{th1,th2}
and the parametrizations of Ref. \cite{KZ,MS}.

From  Figs. \ref{fg1} and \ref{fg2}, we realize that
  limits (\ref{a3}) and   (\ref{a4}),
essentially,
determine the energy for $k_F|a|< 1$ and
 $k_F|a|> 10$.
The correct energy over the entire crossover  should be a smooth interpolation
between these limits.
This is what has been done  to obtain the present results for $a<0$ and
$a>0$ in very good agreement with the  accurate GFMC calculations
\cite{th1,th2}
and  asymptotic limits.

To study the collective and bulk properties of the superfluid Fermi gas,
we  write convenient mean-field equations. They are,
essentially, the Schr\"odinger equation with a nonlinear term
equal to the bulk chemical potential  $\mu(n,a)=\partial{\cal
E}/\partial
n$, with ${\cal E}\equiv (E/N) n$ the energy density. A straightforward
calculation with the leading two terms of  energy of Eq. (\ref{a3})
yields  the following bulk chemical potential for $a\to -0$ 
\cite{TH}
\beq
\label{mu1}
\mu(n,a)
=An^{2/3}+{2\pi\hbar^2
a}n/m,
\eeq
to be used
in the nonlinear  Schr\"odinger
equation 
\beq
i \hbar {\partial \over \partial t} \Psi
= \Big[ -{\hbar^2\over 2m^*}\nabla^2_{\bf r} + U({\bf r})
+ \mu(n,a) \Big] \Psi \; ,
\label{bbm}
\eeq
where   $U(\bf r)$
is a trap, 
$({\bf r}, t)$ are space and time
variables, density $n$ is related to order parameter $\Psi$
by $n=|\Psi|^2$, such that $\int n d{\bf r}=N$, and  $m^*\approx 2m$ 
 is the effective mass of pairs.
This equation can also be interpreted to be the Euler-Lagrange equation 
of a time-dependent density functional theory \cite{MS,hu}. Also,  
for a large number of atoms $N$, for a description of collective 
properties  Eq. (\ref{mu1}) is completely 
equivalent to the quantum hydrodynamic equations \cite{TH}.  
The first term on the right-hand side of Eq. (\ref{mu1}) is the standard
nonlinearity in the three-dimensional GL equation of a Fermi superfluid
(equal to Fermi energy)
with the last term appearing due to atomic interaction. This last term
is reduced by a factor of 2 compared to the GP nonlinearity for bosons
($4\pi\hbar^2an/m$),
as, in the present case,
 half of the atomic interactions, those between spin-parallel
fermions, are inoperative.

 The bulk chemical potential $\mu(n,a)$
corresponding to  BBM  energy (\ref{a2})
\begin{eqnarray}\label{xy}
\mu(n,a)
=An^{2/3}\frac{1-\frac{4(3\pi^2n)^{1/3}}{9\pi}a\frac{2+3\xi}{1-\xi}
+\frac{100(3\pi^2n)^{2/3}\xi}
{81\pi^2(1-\xi)^2}a^2}{\left(1-\frac{10
(3\pi^2n)^{1/3}}{9\pi(1-\xi)}
a\right)^2}
\end{eqnarray}
is  to be used in Eq. (\ref{bbm}).  A
simpler expression for bulk chemical potential
can be obtained  if we recall that, in the
unitarity limit ($a\to -\infty$), the energy is
given by $E/N= 3E_F\xi /5$
corresponding to a
bulk chemical potential \cite{baker}
\beq
\label{mu2}
\mu(n,a)={\hbar^2}(3\pi^2n)^{2/3}\xi/(2m).
\eeq
In the full crossover problem, combining the limiting values
(\ref{mu1})
and (\ref{mu2}), we suggest the following bulk chemical potential in
 BBM  valid for all negative $a$
\begin{eqnarray}
\mu(n,a)& =&
 An^{2/3}\left(1 +\frac{\frac{4\pi}{(3\pi^2)^{2/3}} a
n^{1/3}}
{1-\frac{4\pi}{(1-\xi)(3\pi^2)^{2/3}} a
n^{1/3}} \right),
\label{mu3} 
\end{eqnarray}
to be used in Eq. (\ref{bbm}). This is the simplest minimal
bulk chemical potential consistent with limits (\ref{mu1})
and (\ref{mu2}). Expression (\ref{xy}) also satisfies these limits
and is consistent with the accurate GFMC calculations.
We have checked through  numerical calculation that expressions 
(\ref{xy}) and (\ref{mu3}) agree to each other
within an error of less than 1$\%$.
We suggest here the GL-type BBM Eqs.
(\ref{bbm}) and (\ref{mu3}) including beyond
mean-field corrections valid
for $a<0$ and specially in the unitarity limit $a\to -\infty$.

\begin{figure}[tbp] \begin{center}
{\includegraphics[width=.7\linewidth]{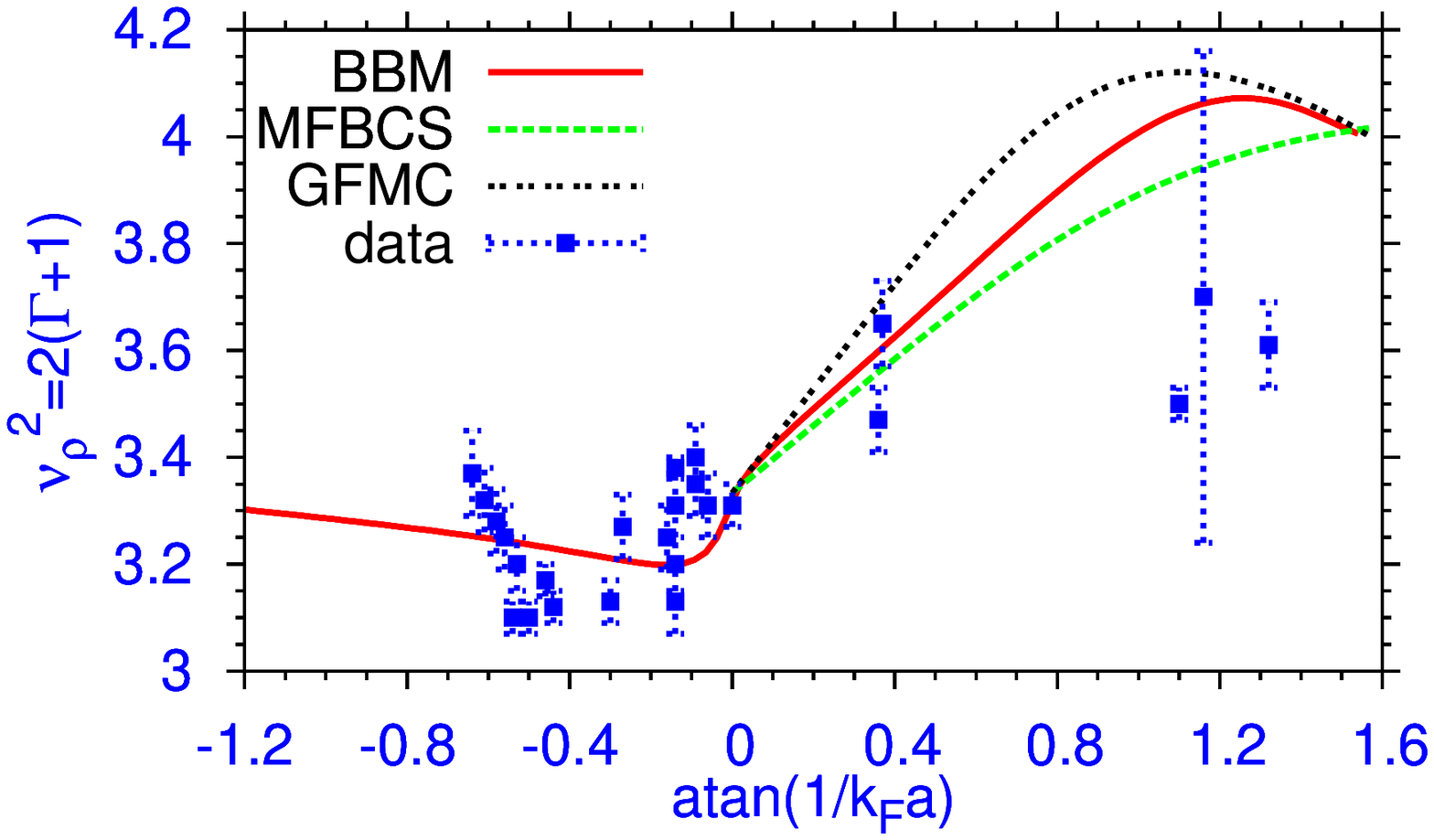}}
{\includegraphics[width=.7\linewidth]{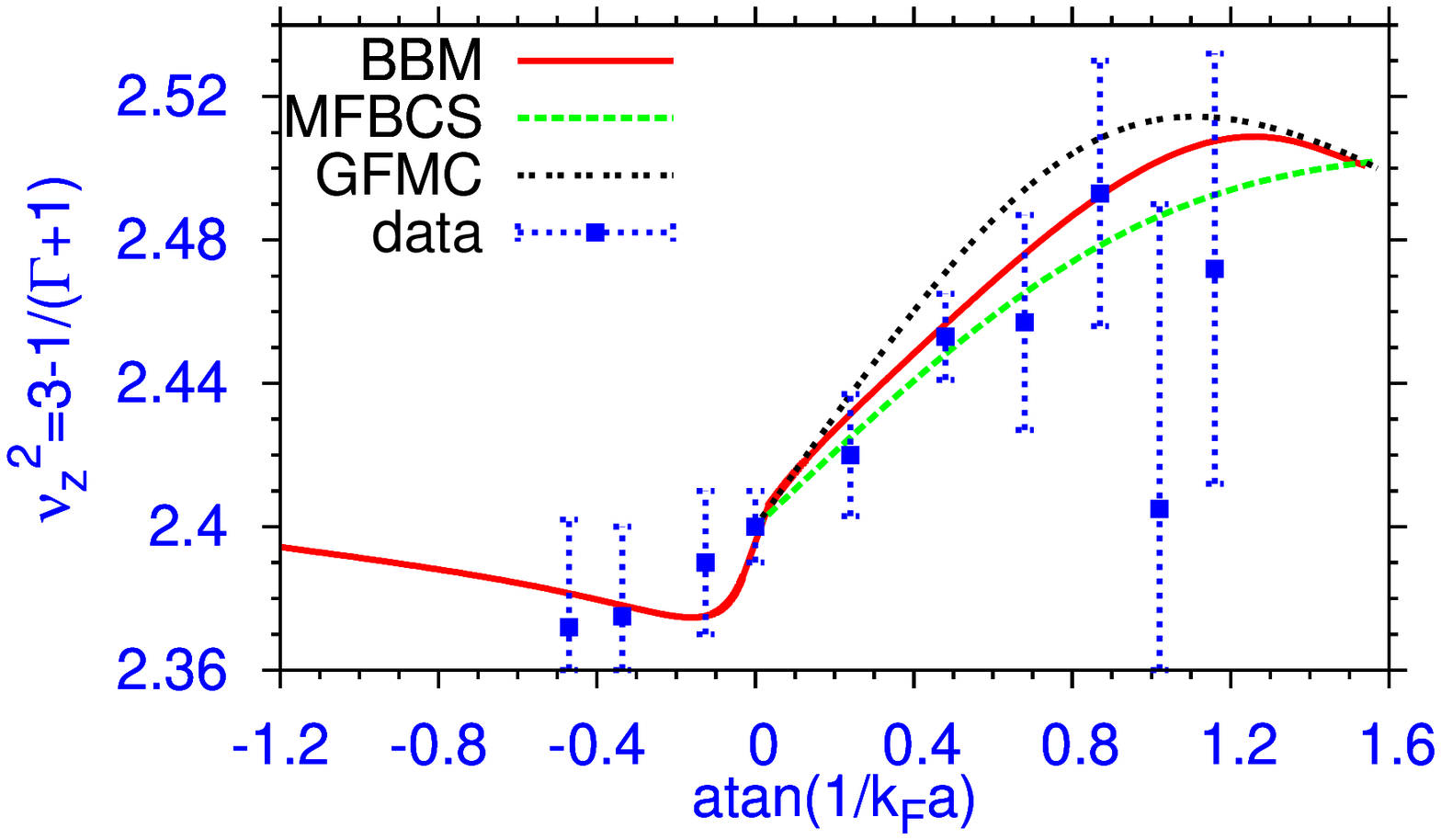}} \end{center}
\caption{(Color online) Square of the radial and axial frequencies (a)
$\nu_\rho^2$ and (b) $\nu_z^2$ in a cigar-shaped trap vs. $k_Fa$, for
BBM (\ref{mu5}), GFMC, MFBCS as quoted in \cite{str}. Data for radial
frequency taken from Refs. \cite{excross2,MS} and those for axial
frequency taken from Refs. \cite{excross1,com,str}. 
}
\label{fg3}
\end{figure}

For $a>0$,
from Eq. (\ref{dimer}), in the small $a'$ limit, the leading terms in
energy density can be written as
\beqa
{\cal E}'=
\frac{2\pi\hbar^2n'^2a'}{m'}+\frac{256\sqrt \pi}{15} \frac{\hbar^2
}{m'}(n'a')^{5/2}.\label{db}
\eeqa
This result, applicable to dimers, is written in terms
of dimer
variables denoted by prime.
and 
is  obtained for a uniform hard-sphere Bose
gas  (here composite bosonic dimers)
in a perturbation calculation for small $n'a'^3$ \cite{polls}.
This leads to a bulk chemical potential 
\cite{TH,lhy}
\beqa
\mu'(n',a')= \frac{4  \pi \hbar^2a'n'}{m'}+128  \sqrt \pi
\frac{\hbar^2a'^{5/2}n'^{3/2}}{3m'}\label{mu4}
\eeqa
in the
mean-field
Eq. (\ref{bbm}), however, now with mass, trap, scattering length,
particle number etc.
appropriate for dimers.
We have recovered in Eq. (\ref{mu4}) the
proper beyond mean-field
generalization of the GP equation  for small $a$ \cite{polls}.
In the large $a$ limit, from Eq.
(\ref{mu2}),   the leading term in
the bulk chemical potential
 is given by
 \cite{TH}
\beq\label{pt}
\mu'(n',a')=\xi \hbar^2(6\pi^2)^{2/3}n'^{2/3}/m'.
\eeq
Equations. (\ref{mu4}) and (\ref{pt}) can be
combined to form the
following  chemical potential
valid for small and large $a$  
\beq
\mu'(n',a')=
\frac{\frac{4\pi\hbar^2 a'n'}{m'}\left(1+\frac{64}{3\sqrt
\pi}a'^{3/2}\sqrt {n'}
   \right)}{1+
\frac{32}{3\sqrt{\pi}}a'^{3/2}\sqrt{n'}
+\frac{64}{3\sqrt
\pi}\frac{2\pi}{(6\pi^2)^{2/3}\xi}
a'^{5/2}
{n'}^{5/6} }.\label{mu5}
\eeq
This bulk chemical potential has been constructed to satisfy
Eq. (\ref{pt})
for large $a'$ and Eq. (\ref{mu4}) for  small $a'$. (After a simple
algebra it can be shown that Eq. (\ref{mu5}) satisfies limit (\ref{a4})
for large $a'$ with $\zeta \approx 1$.)
Expression (\ref{mu5}) is much simpler than that obtained directly from
Eq.
(\ref{dimer2}).  Equations (\ref{bbm}) and (\ref{mu5})
are the
present GP-type BBM equations for the bosonic dimers including beyond
mean-field corrections valid for $a>0$ and especially in the unitarity
limit.
For $\xi=0.44$,
Eq. (\ref{mu5}) produces the following unitarity limit for
composite
bosonic dimers  $\mu'(n',a')= \kappa \hbar^2n'^{2/3}/m', \kappa \approx
7.$
For fundamental
bosons a similar relation is obtained  with a different
coefficient $\kappa$
\cite{cowell}.
Equations  (\ref{bbm}) and (\ref{mu5}) with a different
numerical coefficient $\xi$ appropriate for the study of a
fundamental-boson
superfluid has recently been suggested \cite{as}.

Next we subject bulk chemical potential (\ref{mu5}) to the  stringent 
test
 by calculating the radial and axial frequencies $\nu_\rho$ 
and $\nu_z$ of
collective oscillation in a cigar shaped trap, where ${\bf r}\equiv 
(\rho,z)$ are the radial ($\rho$)  and axial ($z$) coordinates. Cozzini 
and Stringari
\cite{TH,CS} showed that for a power-law politropic
dependence of bulk chemical 
potential on
density $\mu'\propto n'^\Gamma$, $\nu_\rho$ and $\nu_z$ 
are 
given by   $\nu_\rho ^2=2(\Gamma+1)$  \cite{MS}
  $\nu_z^2=3-1/\Gamma(\Gamma+1)$, respectively,  with
$\Gamma=\frac{n'}{\mu'}\frac{\partial \mu'}{\partial n'}$. 
To test the
BBM,
we
plot, in Fig. \ref{fg3} (a) and (b), $\nu_\rho^2$ and $\nu_z^2$ vs. 
atan$(1/k_Fa)$, respectively,  
and compare them  with the
GFMC and  MFBCS calculations, and experimental data of Refs. 
\cite{excross1,excross2}, as quoted in Refs. 
\cite{MS,str,com}.   
The end points of the present BBM 
plot in
Fig. \ref{fg3} are determined by the value of $\Gamma= 2/3$ and 1 at 
$a=+\infty$ and 0, respectively. 
However, as $\nu^2$ is related
to the
derivative ${\partial \mu'}/{\partial n'}$,   bulk chemical 
potential (\ref{mu5}) provides
a good fit of the derivative of the $\mu'-n'$ curve to
theoretical models \cite{2,str}  and experiment 
\cite{excross1,str}  in 
addition to the 
$\mu'-n'$ curve as can be
seen from Fig. \ref{fg3}. In this figure the difference between
different curves is small (order of $2\%$ to $3\%$) and by
slightly altering the constants in Eq. (\ref{mu5}) we can
obtain a result close to either  the GFMC or   MFBCS plot. We did 
not do so and the present result is obtained only from asymptotic
condition without any fitting.

\begin{figure}[tbp]
\begin{center}
{\includegraphics[width=.7\linewidth]{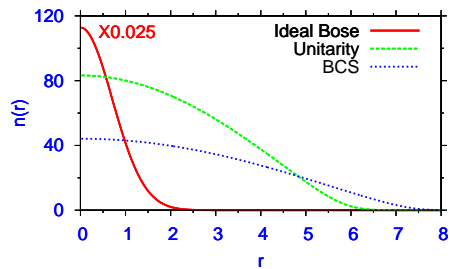}}
\end{center}
\caption{(Color online) Density $n(r)$ vs.  $r$ in
units of oscillator lingth ($l=\sqrt{\hbar/m \omega}$)
in BCS and Bose regimes normalized to $4\pi\int n(r)r^2 dr=N=2000$:
Unitarity  $|a|\to  \infty$;  Ideal Bose
$a\to +0$;   BCS
$a\to -0$. }
\label{fg4}
\end{figure}

Finally,  we solve, numerically, for a
spherical harmonic trap,
Eqs. (\ref{bbm}) and (\ref{mu3})  for fermions ($a<0$)
 and  Eqs. (\ref{bbm}) and  (\ref{mu5})
for dimer bosons
($a>0$)
by the method of imaginary time propagation after 
discretizing it with the semi-implicit Crank-Nicholson rule.  
We employ
length scale $l=\sqrt{\hbar/m \omega}$ and time scale 
($\tau=\omega^{-1}$)
with  $\omega$
the angular frequency of  trap,
and
2000 fermionic atoms (1000 dimers). In 
numerical simulation we use  $l=0.025$ and $\tau =0.001$. 
The calculated densities
plotted in Fig. \ref{fg4} show interesting behavior.
The rms radius $r_{rms}$ in different regimes obey  inequalities
$r_{rms}^{a\to -0}> r_{rms}^{|a|\to \infty}>
r_{rms}^{a\to +0}$ with numerical values $4.74\mu$m,
$3.85 \mu$m,
and
$1.25 \mu$m, respectively. The size depends on the respective
nonlinearity,  increasing as the nonlinearity increases.
In the BCS limit ($a\to -0$)
the Fermi gas extends to a greater distance than at
unitarity due to an  increased repulsion.
The ideal  dimer Bose gas with  $a'=0$ has the most compact
structure due to no repulsion.

To conclude, we proposed and solved numerically a 
quasi-analytic  nonlinear 
Eq.  (\ref{bbm})
for  Bose to BCS crossover of a  dilute Fermi gas with beyond 
mean-field correction so that it is valid in the unitarity region with 
divergent scattering length $a$.  This model produces the known analytic 
behavior of the energy and bulk chemical potential of the system 
(dependence on scattering length)
in the  
BCS ($a\to -0$), Bose ($a\to +0$), and  unitarity 
$a\to \pm\infty$ limits.
For $a<0$ (Fermi regime), the equations
are GL-type Eqs. (\ref{bbm}) and (\ref{mu3}), and for $a>0$ (dimer 
Bose regime), they are
the
GP-type Eqs. (\ref{bbm})
and (\ref{mu5}). The calculated radial and axial frequencies of 
collective breathing oscillation of a system in a cigar-shaped trap are 
found to be in good agreement with experiment and GFMC and MFBCS 
calculations. 

We  thank Luca Salasnich for valuable discussion, Stefano Giorgini for
additional results \cite{th2} and FAPESP
and CNPq (Brazil) for partial support.

\end{document}